\documentclass[aps,prl,twocolumn,superscriptaddress,showpacs]{revtex4-1}
\usepackage{amsmath,amssymb}
\usepackage{graphicx}
\usepackage{epsfig}
\usepackage{color}
\usepackage{rotating}
\usepackage{bm}
\usepackage{setspace}

\begin{document}
\title{Stoichiometry and defect superstructures in epitaxial FeSe films on SrTiO$_3$}
\author{Xue-Qing Yu}
\author{Ming-Qiang Ren}
\author{Yi-Min Zhang}
\author{Jia-Qi Fan}
\author{Sha Han}
\affiliation{State Key Laboratory of Low-Dimensional Quantum Physics, Department of Physics, Tsinghua University, Beijing 100084, China}
\author{\\Can-Li Song}
\email[]{clsong07@mail.tsinghua.edu.cn}
\author{Xu-Cun Ma}
\email[]{xucunma@mail.tsinghua.edu.cn}
\affiliation{State Key Laboratory of Low-Dimensional Quantum Physics, Department of Physics, Tsinghua University, Beijing 100084, China}
\affiliation{Frontier Science Center for Quantum Information, Beijing 100084, China}
\author{Qi-Kun Xue}
\email[]{qkxue@mail.tsinghua.edu.cn}
\affiliation{State Key Laboratory of Low-Dimensional Quantum Physics, Department of Physics, Tsinghua University, Beijing 100084, China}
\affiliation{Frontier Science Center for Quantum Information, Beijing 100084, China}
\affiliation{Beijing Academy of Quantum Information Sciences, Beijing 100193, China}

\begin{abstract}
Cryogenic scanning tunneling microscopy is employed to investigate the stoichiometry and defects of epitaxial FeSe thin films on SrTiO$_3$(001) substrates under various post-growth annealing conditions. Low-temperature annealing with an excess supply of Se leads to formation of Fe vacancies and superstructures, accompanied by a superconductivity (metal)-to-insulator transition in FeSe films. By contrast, high-temperature annealing could eliminate the Fe vacancies and superstructures, and thus recover the high-temperature superconducting phase of monolayer FeSe films. We also observe multilayer FeSe during low-temperature annealing, which is revealed to link with Fe vacancy formation and adatom migration. Our results document very special roles of film stoichiometry and help unravel several controversies in the properties of monolayer FeSe films.
\end{abstract}

\maketitle
\begin {spacing}{1.005}
High-temperature ($T_\textrm{c}$) superconductivity in a number of FeSe-related compounds has attracted worldwide attention in the community of superconductors \cite{qing2012interface, Guo2010superconductivity, burrard2013enhancement, lu2014coexistence, miyata2015high, Lei2016evolution, observation2016song, shiogai2016electric, Shahi2018high}, with a particular focus on monolayer (ML) FeSe epitaxial films grown on SrTiO$_3$ substrates by molecular beam epitaxy (MBE) \cite{wen2014direct, he2013phase, tan2013interface,ge2014superconductivity,xiang2012high, lee2014interfacial, fan2015plain, liu2017topological, zhang2017origin, zhao2018direct,song2019evidence}. This seems understandable because the structurally simple FeSe/SrTiO$_3$ system exhibits an unexpectedly high $T_\textrm{c}$ ($\sim$ 65 K or even higher) \cite{qing2012interface, wen2014direct, he2013phase, tan2013interface,ge2014superconductivity} and serves as an ideal platform to unravel the mystery of unconventional superconductivity in iron-based compounds. Since the discovery, much has been learned about electron transfer from the substrates to monolayer FeSe that significantly enhances $T_\textrm{c}$ \cite{Guo2010superconductivity, burrard2013enhancement, lu2014coexistence, miyata2015high, Lei2016evolution, observation2016song, shiogai2016electric, Shahi2018high, he2013phase, tan2013interface, zhang2017origin, zhao2018direct, song2019evidence}, but unsolved issues including the samples themselves remain. Monolayer FeSe films prepared in different methods display a great diversity of $T_\textrm{c}$ and the superconducting energy gap \cite{huang2017monolayer}. This situation becomes especially prominent when comparing \textit{in-situ} and \textit{ex-situ} measurements \cite{qing2012interface, wen2014direct, he2013phase, tan2013interface, ge2014superconductivity}. The actual roles played by the capping layer (e.g.\ amorphous Se) required for \textit{ex-situ} measurements have been little investigated \cite{wen2014direct, he2013phase}. In order to achieve the high-$T_\textrm{c}$ superconductivity in monolayer FeSe films, the Se capping layer has to be removed via a high-temperature annealing, during which the identified phase transition from a low-doping normal state to superconductivity is controversial in nature \cite{he2013phase, Berlijn2014doping, hu2019insulating}.

On the other hand, intrinsic defects unavoidably occur in epitaxial FeSe films and have been found to severely influence superconductivity \cite{chen2014fe, Jiao2017impurity, Chen2017absence, Liu2018extensive, Hanzawa2019insulator, Bu2019study}. For example, the dumbbell-like defects and their ordering, which have been assigned to Fe vacancies \cite{huang2016dumbbell, Fang2016tunable}, are very detrimental to superconductivity \cite{chen2014fe,Jiao2017impurity,Fang2016tunable}. With regard to film stoichiometry, a recent scanning tunneling microscopy (STM) study argued an unexpected existence of at least 20$\%$ excess Fe in the superconducting FeSe monolayer films by annealing Se-covered FeSe monolayer on SrTiO$_3$ and estimating the amount of newly formed multilayer FeSe islands \cite{Hu2018identification}. This challenges the well-established 1:1 stoichiometry of FeSe for attaining superconductivity in both bulk counterpart and epitaxial films on graphene substrates \cite{williams2009stoichiometry, song2011direct, song2011molecular}. In this study, we investigate systematically the morphology, defects and electronic structure of epitaxial FeSe thin films on SrTiO$_3$(001) by using various post-growth annealing conditions, aiming to clarify the roles of stoichiometry.

Our experiments were conducted on two ultrahigh vacuum (UHV) STM systems (Unisoku), every of which is connected to an MBE chambers for \textit{in-situ} sample preparation. The base pressure of all UHV chambers is better than 2.0 $\times$ 10$^{-10}$ Torr. Nb-doped SrTiO$_3$(001) substrates (0.05 wt$\%$) were degassed at 600 $^\textrm{o}$C, and subsequently annealed at 1250 $^\textrm{o}$C in MBE for 20 minutes to obtain a clean (2 $\times$ 2) surface. High-purity Fe (99.995$\%$) and Se (99.9999$\%$) sources evaporated from the respective Kundsen cells were co-deposited onto the substrates under Se-rich condition. Direct current heating was used to heat the substrates, while the substrate temperature ($T_{\textrm{sub}}$) was measured by a pyrometer. All STM topographies and spectroscopic data were acquired at 4.2 K using polycrystalline PtIr tips, which were cleaned by electron-beam heating and calibrated on MBE-grown Ag/Si(111) films prior to the measurements. Tunneling conductance spectra were collected using a standard lock-in technique with a bias modulation at 931 Hz.
\end {spacing}

\begin{figure}[t]
\includegraphics[width=1\columnwidth]{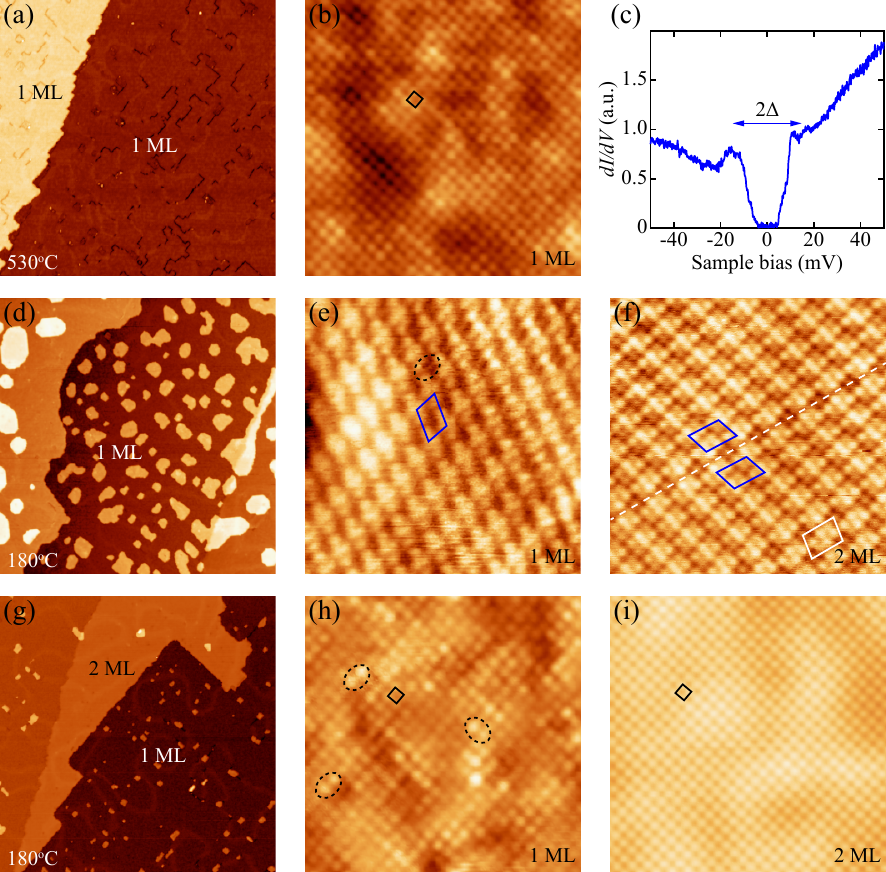}
\caption{(color online) (a) Morphology of monolayer FeSe films (300 nm $\times$ 300 nm, $V$ = 3.0 V, $I$ = 30 pA) on SrTiO$_3$(001) upon annealing at 530 $^\textrm{o}$C. The step-terrace feature stems from a miscut of the underlying SrTiO$_3$ substrates throughout. (b) Atomically resolved STM image (10 nm $\times$ 10 nm, $V$ = 50 mV, $I$ = 100 pA) of monolayer FeSe film in (a). The black square marks the (1 $\times$ 1) unit cell of Se-terminated FeSe(001) surface. (c) Representative \textit{dI/dV }spectrum presenting a superconducting gap of $\Delta$ $\sim$ 15 meV on the high-temperature annealed FeSe films in (a). The setpoint is stabilized at $V$ = 50 mV and $I$ = 100 pA. (d) STM topography (300 nm $\times$ 300 nm, $V$ = 4.0 V, $I$ = 25 pA) showing multilayer FeSe islands after annealing Se-covered monolayer FeSe films at 180 $^\textrm{o}$C. (e, f) STM topographic images (10 nm $\times$ 10 nm, $V$ = 1.8 V, $I$ = 100 pA) of monolayer and bilayer FeSe surfaces in (d), respectively. The blue and white parallelograms correspond to the unit cells of 2 $\times$ $\sqrt{10}$ and $\sqrt{5}$ $\times$ $\sqrt{10}$ surface superstructures, respectively. (g) STM topography (300 nm $\times$ 300 nm, $V$ = 4.0 V, $I$ = 30 pA) showing the coexistence of 1 ML and 2 ML FeSe after annealing monolayer FeSe films in MBE chamber at 180 $^\textrm{o}$C. (h, i) STM topographic images (10 nm $\times$ 10 nm, $V$ = 0.5 V, $I$ = 100 pA) of monolayer and bilayer FeSe surfaces in (g), respectively.
}
\end{figure}

We begin with superconducting monolayer FeSe films [Figs.\ 1(a)-1(c)], which are all prepared by depositing $\sim$ 1.5 ML FeSe on the as-cleaned SrTiO$_3$ substrates at $T_{\textrm{sub}}$ = 450 $^\textrm{o}$C, followed by annealing at a higher temperature of $\sim$ 530 $^\textrm{o}$C. Such a post-growth annealing was found to eliminate effectively extra FeSe islands of $\sim$ 0.5 ML [Fig.\ 1(a)] that was interpreted as decomposition of FeSe \cite{zhang2017origin}. Amorphous Se of $\sim$ 10 ML (here 1 ML is defined as the area density of Se atoms in single-layer FeSe film, $\sim$ 1.4 $\times 10^{15}/cm^2$) was subsequently deposited on the superconducting monolayer FeSe films, followed by annealing at lower $T_{\textrm{sub}}$ = 180  $^\textrm{o}$C for 2 hours. A representative STM topographic image of the annealed films is depicted in Fig.\ 1(d). Apparently, multilayer FeSe islands with a coverage up to 0.45 $\pm$ 0.14 ML are formed, analogous to the previous study \cite{Hu2018identification}. More remarkably, a closer examination reveals a dominant 2 $\times$ $\sqrt{10}$ surface superstructure in both 1 ML and 2 ML FeSe films, marked by the blue parallelograms in Figs.\ 1(e) and 1(f). Occasionally, we observe some antiphase domain boundaries. As illustrated by the white dashed lines in Fig. 1(f), the boundary separates two adjacent domains of 2 $\times$ $\sqrt{10}$ phase that shift by one in-plane Se-Se spacing of $a_{\textrm{Se-Se}}$ = 3.78 \AA.

For comparison, we have also prepared another monolayer FeSe films, similar to the one in Fig.\ 1(a), and annealed it directly at 180 $^\textrm{o}$C without prior Se deposition. Interestingly, multilayer FeSe islands develop both along the step edges and on the terraces [Fig.\ 1(g)], despite a relatively smaller coverage of 0.26 $\pm$ 0.03 ML. Another remarkable distinction from the low-temperature annealed FeSe films with an excess supply of Se in Figs.\ 1(d)-(f) is that the FeSe(001) surface remains basically unchanged and exhibits no surface superstructure [Figs.\ 1(h) and 1(i)], except for an increase of dumbbell-like Fe vacancies in number, marked by the black ellipses in Fig.\ 1(h). The populated Fe vacancies were found to completely kill the superconductivity in monolayer FeSe films on SrTiO$_3$ substrates.

In order to unravel the formation mechanism of multilayer FeSe islands, Fe vacancies and 2 $\times$ $\sqrt{10}$ surface superstructure, we conducted progressive annealing of the superstructural FeSe thin films [Figs.\ 1(d)-1(f)] at elevated temperatures, studied the surface structure, and measured the corresponding tunneling conductance spectra [Fig.\ 2]. With increasing $T_{\textrm{sub}}$, the multilayer FeSe islands gradually reduce in population and eventually vanishes at 530 $^\textrm{o}$C \cite{Hu2018identification}, restoring to the superconducting monolayer phase as Fig.\ 1(a). Meanwhile, the 2 $\times$ $\sqrt{10}$ surface superstructure disappears at $\sim$ 250 $^\textrm{o}$C, and the monolayer FeSe films are populated by a considerable of randomly distributed Fe vacancies below $T_{\textrm{sub}}$ = 300 $^\textrm{o}$C. As $T_{\textrm{sub}}$ is further increased, the Fe vacancies are significantly reduced, thereby recovering the Se-terminated FeSe(001) surface [Fig.\ 2(a)]. Moreover, the Fe vacancies are easier to be removed in 2 ML films [Fig.\ 2(a)]. Such a reduction of dumbbell-like Fe vacancies with post-growth annealing bears a striking resemblance to MBE-grown FeSe films on graphitized SiC(0001) substrates \cite{song2011direct}, and is possibly associated with the diffusion of Fe vacancies to the edge of terraces \cite{huang2016dumbbell}.       ¡¡

\begin{figure*}[t]
\includegraphics[width=1.97\columnwidth]{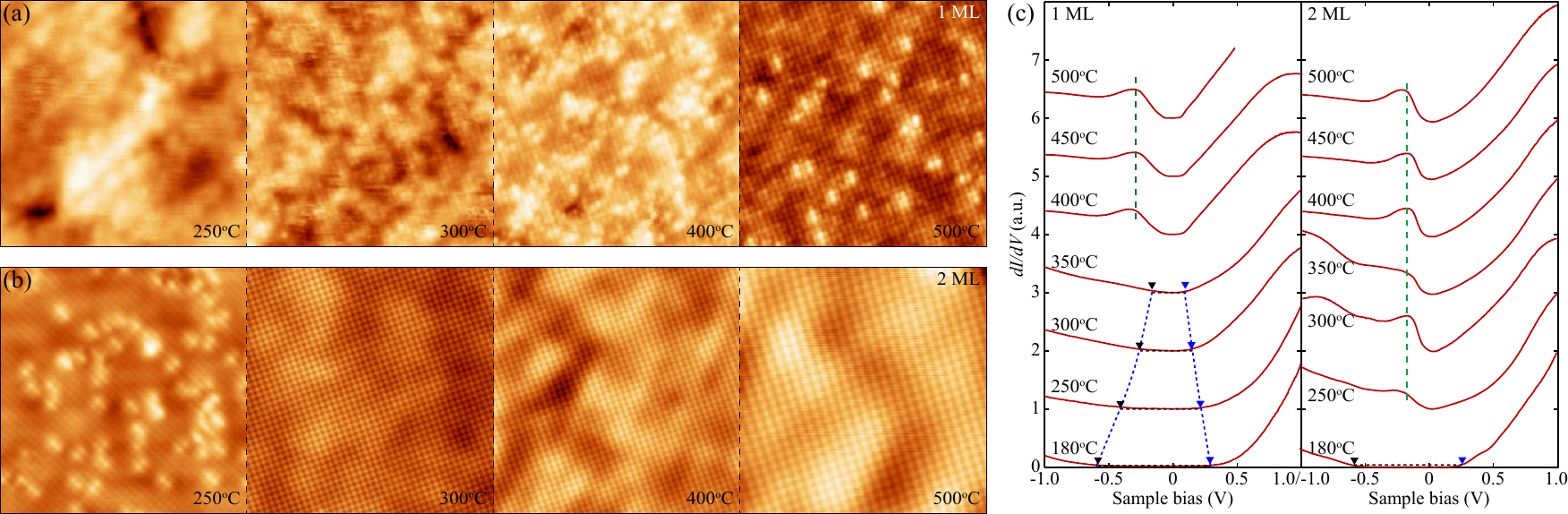}
\caption{(color online) (a, b) Evolution of surface structure with annealing temperature as indicated in monolayer and bilayer FeSe(001)$-$2 $\times$ $\sqrt{10}$ films. The dimension of STM images is 15 nm $\times$ 15 nm except for the one of bilayer FeSe annealed at 500 $^\textrm{o}$C (10 nm $\times$ 10 nm). The imaging conditions are $I$ = 100 pA and $V$ = (a) 0.8 V, 0.5 V, 0.1 V, 0.05 V, (b) 0.1 V, 0.2 V, 0.1 V, 0.2 V from left to right. The duration time is 2 hours for every annealing sequence. (c) Tunneling \textit{dI/dV} spectra as a function of annealing temperature. Black and blue triangles mark the onsets of valance band and conduction band throughout, respectively. Horizontal black dashes denote the band gap size of insulating FeSe films with populated Fe vacancies, while the vertical dashes are guide to eyes. The tunneling spectra are vertically offset for clarify. Setpoint: $V$ = 1.0 V and $I$ = 100 pA.
}
\end{figure*}

\begin{figure*}[t]
\includegraphics[width=1.5\columnwidth]{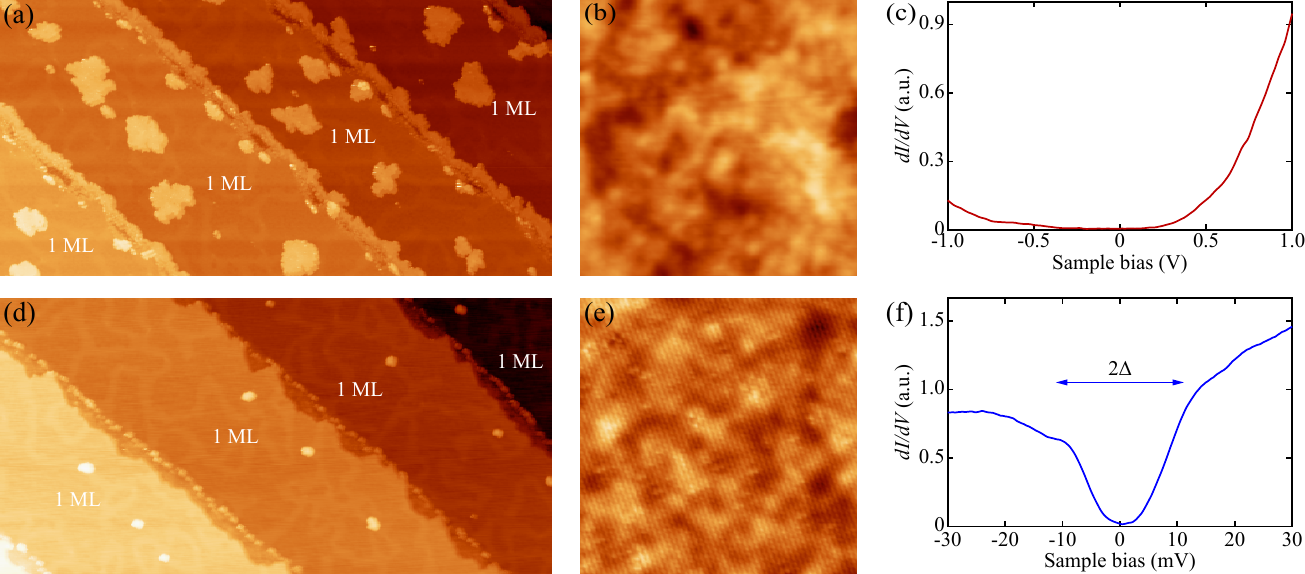}
\caption{(color online) (a) STM topography (480 nm $\times$ 240 nm, $V$ = 3.0 V, $I$ = 10 pA) of superconducting monolayer FeSe films after kept in Se-containing MBE chamber at RT for 16 hours. (b) STM topographic image (30 nm $\times$ 30 nm, $V$ = 3.0 V, $I$ = 10 pA) and (c) typical \textit{dI/dV} spectrum ($V$ = 1.0 V, $I$ = 100 pA) on monolayer FeSe films in (a). (d) Topography (480 nm $\times$ 240 nm, $V$ = 3.0 V, $I$ = 10 pA) of superconducting monolayer FeSe films after kept in a clean UHV chamber at RT for 48 hours. (e) STM topographic image (30 nm $\times$ 30 nm, $V$ = 50 mV, $I$ = 50 pA) and (c) superconducting gap ($V$ = 30 mV, $I$ = 100 pA) on monolayer FeSe films in (d). Note that the weak meandering patterns in (a) and (d) correspond to the domain boundaries of FeSe films.
}
\end{figure*}

Figure 2(c) plots the spatially averaged tunneling conductance \textit{dI/dV} spectra measured on both 1 ML and 2 ML FeSe films as a function of annealing temperature. At $T_{\textrm{sub}}$ = 180 $^\textrm{o}$C, the superstructural 2 $\times$ $\sqrt{10}$ surfaces are insulating with a band gap of up to 0.85 eV. As $T_{\textrm{sub}}$ is elevated, the insulating gap reduces in magnitude and the tunneling \textit{dI/dV} spectra recover eventually the characteristic features of high-temperature annealed FeSe films [the top curves in Fig.\ 2(c)] \cite{Ding2016high}, leading to an insulator-to-superconductivity (metal) transition in 1 ML (2 ML) FeSe films. Apparently, the transition is more abrupt for 2 ML FeSe, echoing  the easier removal of Fe vacancies upon post-growth annealing [Figs.\ 2(a) and 2(b)]. Note that the gap-like conductance spectra in 1 ML FeSe at $T_{\textrm{sub}} \geq$ 400 $^\textrm{o}$C as seen by STM are caused by the significantly small conductance from the electron pockets around $M$ points of the Brillouin zone due to tunneling matrix effect, rather than indicating a real band gap. Thus the phase transition from a low-doping normal/insulating state to superconductivity upon annealing is primarily triggered by the removal of Fe vacancies. The recently observed insulating phase in monolayer FeSe films \cite{hu2019insulating} is essentially off-stoichiometric Fe$_{1-x}$Se compound with a considerable amount of Fe vacancies. Here the Fe vacancies give rise to extra holes and reduce the effective electron carriers for superconductivity, which may be a killer of superconductivity in FeSe.

Now the cause of multilayer FeSe islands upon annealing at 180 $^\textrm{o}$C becomes apparent. Firstly, the formation of Fe-vacancy superstructures provides excess Fe atoms to grow multilayer FeSe [Figs.\ 1(d)-1(f)]. Such argument has been convincingly corroborated in Fig.\ 3. By annealing monolayer FeSe films in Se-containing MBE chamber at room temperature (RT), we also find multilayer FeSe islands of 0.21 $\pm$ 0.02 ML [Fig.\ 3(a)]. Due to the random distribution of Fe vacancies, the FeSe surface gets disordered and insulating [Figs.\ 3(b) and 3(c)], behaving similarly as the Fe-deficient monolayer FeSe films in Figs.\ 2(a) and 2(c). This shows that some Fe atoms have migrated out of the monolayer FeSe films and formed multilayer FeSe, given the fact that there exist Se residues in the MBE chamber. Actually, when a clean UHV chamber free of Se was used to keep the sample, no multilayer FeSe islands were formed [Fig. 3(d)] and the pristine FeSe(001)$-$1 $\times$ 1 surface remains superconductive [Figs.\ 3(e) and 3(f)].

Note that every Fe vacancy brightens two adjacent Se atoms at the top layer of FeSe and leads to the dumbbell-shaped feature. Therefore, the stoichiometry of 2 $\times$ $\sqrt{10}$ superstructure can be readily determined as Fe$_5$Se$_6$, because the superstructure unit cell contains four bright Se atoms and thus two Fe vacancies in Figs.\ 1(e) and 1(f). It should be also noted that other Fe-vacancy superstructures coexist on the annealed FeSe films in Fig.\ 1(d), such as $\sqrt{5}$ $\times$ $\sqrt{10}$ (Fe$_{11}$Se$_{14}$, illustrated in the lower right corner of Fig.\ 1f) \cite{li2014molecular} and $\sqrt{5}$ $\times$ $\sqrt{5}$ (Fe$_4$Se$_5$) [Fig.\ 4(a)]. These Fe-vacancy superstructures contribute to $\sim$ 0.2 ML FeSe, which explain well the formed multilayer FeSe islands in Fig.\ 3(a).

\begin{figure}[b]
\includegraphics[width=0.75\columnwidth]{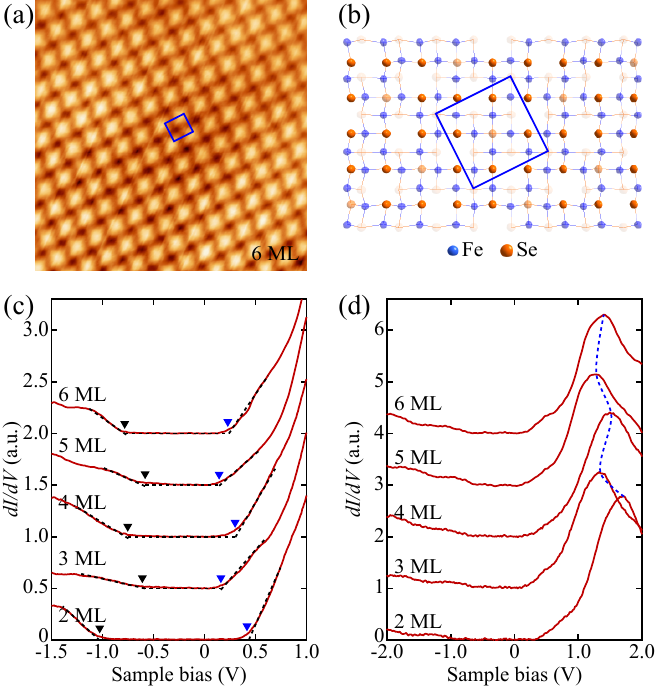}
\caption{(color online) (a) High-resolution STM topography (10 nm $\times$ 10 nm, $V$ = 0.6 V, $I$ = 50 pA) of superstructural $\sqrt{5}$ $\times$ $\sqrt{5}$ (blue square) FeSe films with a thickness of 6 ML. (b) Sketched lattice model of Fe-vacancy Fe$_4$Se$_5$ phase. Note that the Se atoms below the Fe plane are marked by the orange spheres with light opacity. (c) Spatially-averaged \textit{dI/dV} spectra ($V$ = 1.0 V, $I$ = 100 pA) of FeSe(001)$-$$\sqrt{5}$ $\times$ $\sqrt{5}$  films at varied thickness. (d) Tunneling spectra over a wide energy range ($V$ = 2.0 V, $I$ = 100 pA) showing a characteristic conductance peak in the unoccupied states. The blue dashed line is guide to the eyes.
}
\end{figure}

Only the scenario of Fe vacancies and superstructures fails to account for the formed multilayer FeSe up to 0.45 $\pm$ 0.14 ML in Fig.\ 1(d) and 0.26 $\pm$ 0.03 ML in Fig.\ 1(g), especially considering the small population of Se vacancies in Fig.\ 1(g). Here we suggest a dynamical atom migration that also contributes to the formation of multilayer FeSe during the post-growth annealing process. At elevated $T_{\textrm{sub}}$, long-range Fe atom migration may occur, rendering the invisibility of multilayer FeSe due to the limited field of view of STM. Conversely, during the low-temperature annealing, the atom migration and possible reaction with the supplied Se are responsible for the reentrance of multilayer FeSe islands. Here the atom migration might be distinctively driven by the thermal and atom density gradients during the high- and low-temperature annealing, respectively. The atom migration scenario, together with the Fe vacancies and superstructures, might help explain the mysterious FeSe stoichiometry and why monolayer FeSe films are superconductive only when the stoichiometry is nearly 1:1.

Finally, we discuss more on the Fe-vacancy-induced $\sqrt{5}$ $\times$ $\sqrt{5}$ superstructure, which has been previously observed in alkali/alkaline-intercalated iron selenides \cite{wang2011microstructure, li2012phase, Dagotto2013colloquium}. By co-depositing Fe and Se directly at $T_{\textrm{sub}}$ = 230 $^\textrm{o}$C, we obtain superstructural $\sqrt{5}$ $\times$ $\sqrt{5}$ films with thickness of $\geq$ 2 ML [Figs.\ 4(a) and 4(b)]. Plotted in Fig.\ 4(c) are the thickness-dependent tunneling conductance \textit{dI/dV} spectra, all featuring an insulating band gap. The gap and characteristic conductance peak in the unoccupied states [Fig.\ 4(d)] exhibit oscillatory behavior with a period of 2 ML in magnitude (1.44 eV, 0.77 eV, 1.05 eV, 0.73 eV and 1.01 eV from 2 ML to 6 ML) and energy (1.72 eV, 1.34 eV, 1.53 eV, 1.28 eV and 1.40 eV from 2 ML to 6 ML), respectively, the cause of which merits a further theoretical investigation. On the other hand, the gap magnitude of $\sim$ 1.44 eV in 2 ML is greater than 0.85 eV in the 2 $\times$ $\sqrt{10}$ superstructure [Figs.\ 1(c) and 1(d)]. This is most likely caused by more populated Fe vacancies in $\sqrt{5}$ $\times$ $\sqrt{5}$ superstructure. Overall, the gap magnitude of Fe$_4$Se$_5$ films increases when approaching the two-dimensional limit, a consequence of poor electrostatic screening and enhanced quantum confinement.

In summary, we have explicitly revealed by direct STM imaging that the superconductivity property in monolayer FeSe films on SrTiO$_3$ is very sensitive to Fe vacancies. The Fe vacancy formation together with atom migration takes responsibility for the observed multilayer islands in low-temperature annealed FeSe films. Therefore, a fine control of Fe vacancies not only appears to be essential to the surface morphology and structure, but also plays a key role in the superconductivity of FeSe films.

\begin{acknowledgments}
The work is financially supported by the Ministry of Science and Technology of China (Grants No.\ 2017YFA0304600, No.\ 2016YFA0301004, No.\ 2018YFA0305603), the National Natural Science Foundation of China (Grants No.\ 11774192, No.\ 11634007), and in part the Beijing Advanced Innovation Center for Future Chip.
\end{acknowledgments}

%
\end{document}